\documentclass[prx,superscriptaddress,twocolumn]{revtex4-1}
\usepackage{amsmath}    
\usepackage{graphicx}   
\usepackage{verbatim}   
\usepackage{color}      
\usepackage{subfigure}  
\usepackage[hidelinks]{hyperref}

\begin{document}

\title{Beyond linear elasticity: Jammed solids  at finite shear strain and  rate}
\author{Julia Boschan}

\affiliation{Delft University of Technology, Process \& Energy Laboratory, Leeghwaterstraat 39, 2628 CB Delft, The Netherlands;}
\author{Daniel V{\aa}gberg}
\affiliation{Delft University of Technology, Process \& Energy Laboratory, Leeghwaterstraat 39, 2628 CB Delft, The Netherlands;}

\author{Ell\'ak Somfai}
\affiliation{Institute for Solid State Physics and Optics, Wigner Research Center for Physics, Hungarian Academy of Sciences, P.O.~Box 49, H-1525 Budapest, Hungary}
\author{Brian P. Tighe}
\affiliation{Delft University of Technology, Process \& Energy Laboratory, Leeghwaterstraat 39, 2628 CB Delft, The Netherlands;}
\date{\today}

\begin{abstract}
The shear response of soft solids can be modeled with linear elasticity, provided the forcing is slow and weak. Both of these approximations must break down when the material loses rigidity, such as in foams and emulsions at their (un)jamming point -- suggesting that the window of  linear elastic response near jamming is exceedingly narrow. Yet precisely when and how this breakdown occurs remains unclear. To answer these questions, we perform computer simulations of stress relaxation and shear startup experiments in athermal soft sphere packings, the canonical model for jamming. By systematically varying the strain amplitude, strain rate, distance to jamming, and system size, we identify characteristic strain and time scales that quantify how and when the window of linear elasticity closes, and relate these scales to changes in the microscopic contact network. Our findings indicate that the mechanical response of jammed solids are generically nonlinear and rate-dependent on experimentally accessible strain and time scales. 

\end{abstract}

\maketitle

Linear elasticity predicts that when an isotropic solid is sheared, the resulting stress $\sigma$ is directly proportional to the strain $\gamma$ and independent of the strain rate $\dot \gamma$,
\begin{equation}
\sigma = G_0 \gamma  \,,
\label{eqn:linear}
\end{equation}
with a constant shear modulus $G_0$ \cite{landau}.  The constitutive relation (\ref{eqn:linear}) -- a special case of Hooke's law -- 
is a simple, powerful, and widely used model of mechanical response in solids. Yet formally it applies only in the limit of vanishingly slow and weak deformations. In practice materials possess characteristic strain and time scales 
that define a linear elastic ``window'', i.e.~a parameter range wherein Hooke's law is accurate.
Determining the size of this window is especially important in soft solids, where viscous damping and nonlinearity play important roles \cite{barnes}. 
The goal of the present work is to determine when Hooke's law holds, and what eventually replaces it, in packings of soft frictionless spheres close to the (un)jamming transition. 

\begin{figure}[b]
 \includegraphics[width=\columnwidth]{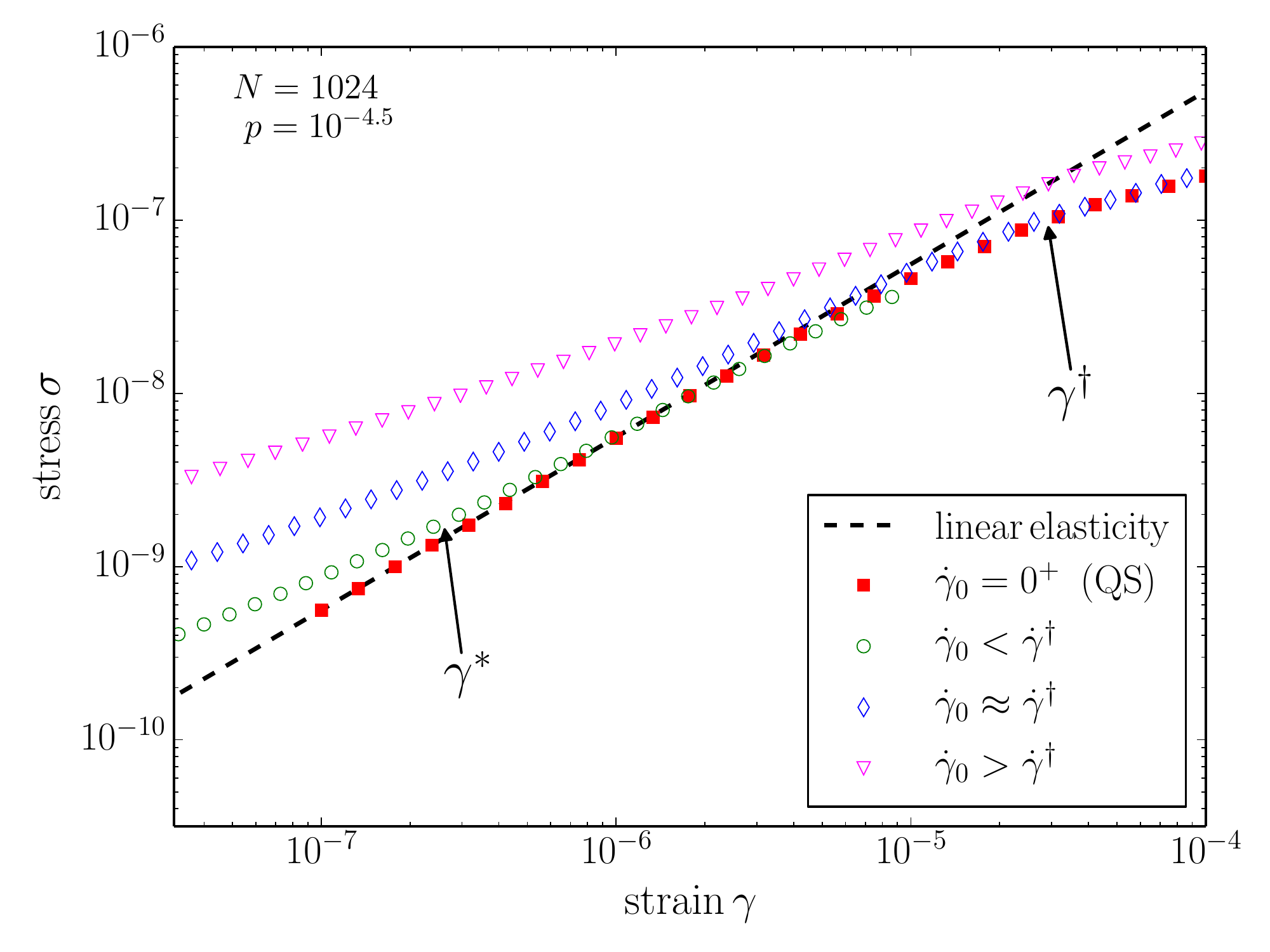}
\caption{Ensemble-averaged stress-strain curves of packings sheared at varying strain rate $\dot \gamma_0$. 
Hooke's law predicts a linear stress-strain curve (dashed line).
The crossover strains $\gamma^*$ and $\gamma^\dag$ are indicated for the data sheared at slow but finite rate $0 < \dot \gamma_0 < \dot \gamma^\dag$ (open circles).
}
\label{fig:window}
\end{figure}

Jammed sphere packings are a widely studied model of emulsions and liquid foams \cite{bolton90,durian95,tighe10c,katgert13} and have close connections to granular media and dense suspensions \cite{hatano09,seth11,franklin15}. 
Linear elastic properties of jammed solids, such as moduli and the vibrational density of states, are by now well understood \cite{vanhecke10,liu10a}. Much less is known about their viscoelastic \cite{head09,hatano09} and especially their nonlinear response \cite{coulais14,otsuki14}. Yet the jamming transition must determine the size of the linear elastic window, because the shear modulus $G_0$  vanishes continuously at the jamming point, where the confining pressure $p$ goes to zero. Indeed, recent studies of oscillatory rheology \cite{tighe11} and shocks \cite{gomez12,ulrich13,vdwildenberg13} have shown that, precisely at the jamming point, {\em any} deformation is effectively fast and strong, and neither viscous effects  nor nonlinearities can be neglected. 

Because elasticity in foams, emulsions, and other amorphous materials results from repulsive contact forces, microstructural rearrangements of the contact network have signatures in the mechanical response. Namely, they lead to nonlinearity and irreversibility in the particle trajectories, and eventually to steady plastic flow \cite{lundberg08,regev13,knowlton14,keim14,keim15,kawasaki15}.
In a series of influential studies, Schreck and co-workers \cite{schreck11,schreck13,schreck13b,schreck14,bertrand14}  recently asked how many contact changes a jammed packing undergoes before its mechanical response ceases to be linear. To answer this question, they studied the onset of mixing between excited vibrational modes in microcanonical ensembles of $N$ particles, and found that trajectories cease to be linear as soon as there is a single rearrangement (made or broken contact) in the contact network. Contact changes occur for perturbation amplitudes that vanish as $1/N$, i.e.~essentially immediately in large systems. Their findings caused the authors to question, if not the formal validity, then at least the usefulness of  linear elasticity in jammed solids -- not just at the jamming point, but anywhere in the jammed phase.

Subsequently, Van Deen et al.~\cite{vandeen14} and Goodrich et al.~\cite{goodrichschreckcomment,goodrich14} argued that the situation is not so dire. They demonstrated that coarse grained properties of jammed solids are far less sensitive to contact changes than are the individual trajectories. Namely, under ensemble averaging the slope of the stress-strain curve is the same before and after the first contact change \cite{vandeen14}, and changes in the density of states are negligible \cite{goodrich14}. These results show that linear elastic constitutive relations remain useful near jamming, but they say nothing about when Hooke's law eventually does break down. 

Recent experiments \cite{coulais14,knowlton14} and simulations \cite{otsuki14,kawasaki15,dagois-bohy14} provide evidence for a two stage yielding process, where packings first soften and only later establish steady shear flow. 
Yet it remains unclear precisely how rate dependence, nonlinearity, and contact changes contribute to the breakdown of linear elasticity. In order to unravel these effects, it is necessary to vary strain, strain rate, pressure, and system size simultaneously and systematically  -- as we do here for the first time.
Using simulations of viscous soft spheres,
 we find that Hooke's law is valid within a surprisingly narrow window bounded by viscous effects at small strain and nonlinear softening at large strain. The size of the linear elastic window displays power law scaling with pressure and correlates with the accumulation of not one, but an extensive number of contact changes.

The basic scenario we identify is illustrated in Fig.~\ref{fig:window}, which presents ensemble-averaged stresses versus strain. Shear is applied via a constant strain rate $\dot \gamma_0$ at fixed volume. We identify three characteristic scales, each of which depend on the initial pressure $p$:
(i) For strains below $\gamma^* \equiv \dot \gamma_0 \tau^*$, where $\tau^*$ is a time scale, viscous stresses are significant and Eq.~(\ref{eqn:linear}) underestimates the stress needed to deform the material. 
This crossover strain  vanishes under quasistatic shear ($\dot \gamma_0 \rightarrow 0$, filled squares). 
(ii) Above a strain $\gamma^\dag$ the material softens and Hooke's law overestimates the stress. This crossover is rate-independent, consistent with plastic effects. 
(iii) For strain rates above a scale $\dot \gamma^\dag$ (triangles), Eq.~(\ref{eqn:linear}) is never accurate and there is no strain interval where the material responds as a linear elastic solid.

\section{Soft spheres: Model and background}
\label{sec:background}
We first introduce the soft sphere model and summarize prior results regarding linear elasticity near jamming.

\subsection{Model}
We perform numerical simulations of the Durian bubble model \cite{durian95}, a mesoscopic model for wet foams and emulsions. The model treats bubbles/droplets as non-Brownian disks that interact via elastic and viscous forces when they overlap. 
Elastic forces are expressed in terms of the overlap $\delta_{ij} = 1 - r_{ij}/{(R_i + R_j)}$, 
where $R_i$ and $R_j$ denote radii and ${\vec r}_{ij}$ points from the center of particle $i$ to the center of $j$. The force is repulsive and acts along the unit vector $\hat r_{ij} = \vec{r}_{ij}/r_{ij}$:
\begin{equation}
   \vec{f}^{\rm el}_{ij} = 
\begin{cases}
    -k(\delta_{ij}) \, \delta_{ij}\, \hat{r}_{ij} \,,  &  \delta_{ij} > 0\\
    {\vec 0},              & \delta_{ij} < 0.
\end{cases}
\label{eq:potential}
\end{equation}
The prefactor $k$ is the contact stiffness, which generally depends on the overlap
\begin{equation}
k = k_0 \, \delta^{\alpha - 2} \,.
\label{eqn:stiffness}
\end{equation} 
Here $k_0$ is a constant and  $\alpha$ is an exponent parameterizing the interaction. In the following we consider harmonic interactions ($\alpha = 2$), which provide a reasonable model for bubbles and  droplets that resist deformation due to surface tension; we also treat  Hertzian interactions ($\alpha = 5/2$), which correspond to elastic spheres.

We perform simulations using two separate numerical methods. The first is a molecular dynamics (MD) algorithm that integrates Newton's laws using the velocity-Verlet scheme. Each disk is assigned a uniform mass $m_i = \pi R_i^2$ proportional to its volume. 
Energy is dissipated by viscous forces that are  proportional to the relative velocity $\Delta {\vec v}^{\,c}_{ij}$ of neighboring particles evaluated at the contact,
\begin{equation}
\vec{f}^{\rm visc}_{ij} =  - \tau_0 \, k(\delta_{ij})\, \Delta {\vec v}^{\,c}_{ij} \,,
\end{equation}
where $\tau_0$ is a microscopic relaxation time. Viscous forces can apply torques, hence particles are allowed to rotate as well as translate.

In addition to MD, we also perform simulations using a nonlinear conjugate gradient (CG) routine \cite{vagberg11}, which keeps the system at a local minimum of the potential energy landscape, which itself changes as the system undergoes shearing. The dynamics are therefore quasistatic, i.e.~the particle trajectories correspond to the limit of vanishing strain rate.

Bubble packings consist of $N = 128$ to $2048$ disks in a 50:50 bidisperse mixture with a 1.4:1 diameter ratio. Shear is implemented via Lees-Edwards ``sliding brick'' boundary conditions.  The stress tensor is given by 
\begin{equation}
\sigma_{\alpha\beta} = \frac{1}{2V} \sum_{ij} f_{ij,\alpha} r_{ij,\beta} - \frac{1}{V} \sum_i m_i v_{i,\alpha} v_{i,\beta} \,,
\label{eqn:stress}
\end{equation}
where $V$ is the volume (area in two dimensions) of the packing, $\vec{f}_{ij}$ is the sum of elastic and viscous contact forces  acting on particle $i$ due to particle $j$, and $\vec{v}_{i}$ is the velocity of particle $i$. Greek indices label components along the Cartesian coordinates $x$ and $y$. 
The confining pressure is $p = - (1/D)(\sigma_{xx} + \sigma_{yy}$), where $D = 2$ is the spatial dimension, while the shear stress is $\sigma = \sigma_{xy} $.
The second term on the righthand side of Eq.~(\ref{eqn:stress}) is a kinetic stress, which is always negligible in the parameter ranges investigated here.
Initial conditions are isotropic with a targeted pressure $p$, prepared using CG and ``shear stabilized'' in the sense of Dagois-Bohy et al.~\cite{dagois-bohy12}, which guarantees that the initial slope of the stress-strain curve is positive. Stresses and times are reported in dimensionless units constructed from $k_0$, $\tau_0$, and the average particle diameter.

\subsection{Distance to jamming}
We use the confining pressure $p$ as a measure of the distance to jamming. The excess volume fraction $\Delta \phi = \phi - \phi_c$ and excess mean contact number $\Delta z = z - z_c$, where $\phi_c$ and $z_c$ refer to the respective values at jamming, are also frequently used for this purpose\cite{vanhecke10,ohern03,katgert10b}. These three alternative order parameters are related via
\begin{equation}
\frac{p}{k} \sim \Delta \phi \sim \Delta z^2 \,.
\label{eqn:order}
\end{equation}
Here $k$ should be understood as a typical value of the contact stiffness in Eq.~(\ref{eqn:stiffness}). The harmonic case ($\alpha = 2$) is straightforward because the contact stiffness is a constant. For other values of $\alpha$, however, $k$ depends on the pressure. As the typical force trivially reflects its bulk counterpart, $f \sim p$,  the contact stiffness scales as
$k  \sim {f}/{\delta} \sim p^{(\alpha - 2)/(\alpha - 1)} $.
In the following, all scaling relations will specify their dependence on $k$ and the time scale $\tau_0$. In the present work $\tau_0$ is independent of the overlap between particles (as in the viscoelastic Hertzian contact problem \cite{ramirez99}), but we include $\tau_0$ because one could imagine a damping coefficient $k \tau_0$ with more general overlap dependence than the form treated here. 

\begin{figure}
  \includegraphics[width=\columnwidth]{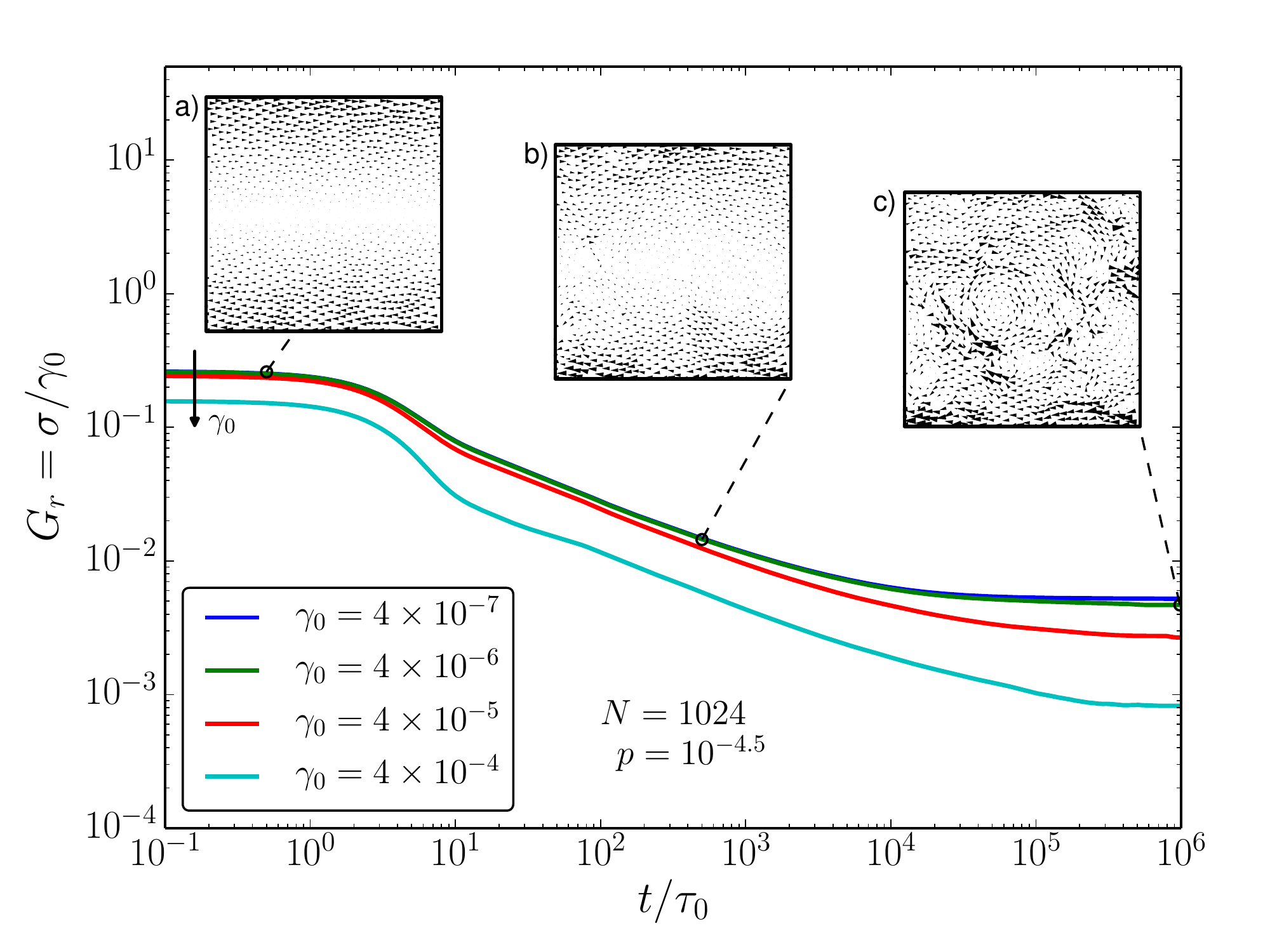}
\caption{The ensemble-averaged relaxation modulus $G_r$ at pressure $p = 10^{-4.5}$ for four values of the strain amplitude $\gamma_0$. 
In all four cases, $G_r$ displays an initial plateau corresponding to affine particle motion (inset a), followed by a power law decay as the particle displacements become increasingly non-affine (b). At long times the stress is fully relaxed and the final particle displacements are strongly non-affine (c). 
}
\label{fig:stressrelax}
\end{figure}

\subsection{Shear modulus and the role of contact changes}
In large systems the linear elastic shear modulus $G_0$ vanishes continuously with pressure, 
\begin{equation}
G_0/k  \sim (p/k)^{\mu} \,,
\label{eqn:G0}
\end{equation}
with $\mu = 1/2$.
Hence jammed solids' shear stiffness can be arbitrarily weak. 
The scaling of $G_0$ has been determined multiple times, both numerically: \cite{ohern03,zhang05,ellenbroek06} and theoretically: \cite{wyartannales,zaccone11,tighe11}; it is verified for our own packings in Fig.~\ref{fig:FS}a, as discussed in Section \ref{sec:relax}. 

There are two standard approaches to determining $G_0$. The first, which we employ, is to numerically impose a small shear strain and relax the packing to its new energy minimum \cite{ohern03,zhang05}. 
In the second approach one writes down the $DN$ equations of motion and linearizes them about a reference state, which results in a matrix equation that can be solved for the response to an infinitesimally weak shear \cite{silbert05,wyart05,ellenbroek06,zaccone11,tighe11,dagois-bohy12}. This latter approach allows access to the zero strain limit, but it is blind to the influence of contact changes.
Van Deen et al.~\cite{vandeen14} verified that the two approaches agree, provided that the strain amplitude is small enough that the packing neither forms new contacts, nor breaks existing ones. They further found that the typical strain at the first contact change depends on pressure and system size as
\begin{equation}
\gamma_{\rm cc}^{(1)}\sim \frac{(p/k)^{1/2}}{N} \,.
\label{eqn:cc}
\end{equation}
Similar to the findings of Schreck et al.~\cite{schreck11}, this scale vanishes in the large system limit, even at finite pressure.

\section{Stress relaxation} 
\label{sec:relax}

We will characterize mechanical response in jammed solids using stress relaxation and flow start-up tests, two standard rheometric tests. In the linear regime they are equivalent to each other and to other common tests, including creep response and oscillatory rheology, as complete knowledge of the results of one test permits calculation of the others \cite{barnes}. This equivalence breaks down once the response becomes nonlinear. 

We employ stress relaxation tests to access the time scale $\tau^*$ over which viscous effects are significant, and we use flow start-up tests to determine the strain scale $\gamma^\dag$ beyond which the stress-strain curve becomes nonlinear. We consider stress relaxation first.

In a stress relaxation test one measures the time-dependent stress $\sigma(t,\gamma_0)$ that develops in a response to a sudden shear strain with amplitude $\gamma_0$, i.e.
\begin{equation}
\gamma(t) = 
\left \lbrace
\begin{array}{cl}
0 & t< 0 \\
\gamma_0 & t \ge 0 \,.
\end{array} \right.
\end{equation} 
The relaxation modulus is
\begin{equation}
G_r(t, \gamma_0) \equiv \frac{\sigma(t,\gamma_0)}{\gamma_0} \,.
\end{equation}
We determine the relaxation modulus by employing the shear protocol of Hatano \cite{hatano09}. A packing's particles and simulation cell are affinely displaced in accordance with a simple shear with amplitude $\gamma_0$. E.g.~for a simple shear in the $\hat x$-direction, the position of a particle $i$ initially at $(x_i, y_i)$ instantaneously becomes $(x_i + \gamma_0  y_i, y_i)$, while the Lees-Edwards boundary conditions are shifted by $\hat \gamma_0 L_y$, where $L_y$ is the height of the simulation cell. Then the particles are allowed to relax to a new mechanical equilibrium while the Lees-Edwards offset is held fixed.

The main panel of Fig.~\ref{fig:stressrelax} illustrates four relaxation moduli of a single packing equilibrated at pressure $p = 10^{-4.5}$ and then sheared with strain amplitudes varying over three decades. All four undergo a relaxation from an initial plateau at short times to a final, lower plateau at long times. The character of the particle motions changes as relaxation progresses in time. While the particle motions immediately after the deformation are affine (Fig.~\ref{fig:stressrelax}a), they become increasingly non-affine as the stresses relax to a new static equilibrium (Fig.~\ref{fig:stressrelax}b,c). 
This non-affine motion is a consequence of slowly relaxing eigenmodes of the packing that become increasingly abundant on approach to jamming \cite{tighe11}. These modes favor sliding motion between contacting particles \cite{ellenbroek06}, reminiscent of zero energy floppy modes \cite{alexander}, and play an important role in theoretical descriptions of mechanical response near jamming \cite{wyart05,wyartannales,maloney06b,zaccone11,tighe11}.

For sufficiently small strain amplitudes, linear response is obtained and any dependence of the relaxation modulus on $\gamma_0$ is sub-dominant. The near-perfect overlap of the moduli for the two smaller strain amplitudes Fig.~\ref{fig:stressrelax} indicates that they reside in the linear regime. The long-time plateau is then equal to the linear elastic modulus $G_0$.
In practice there is a crossover time scale $\tau^*$ such that for longer times $t \gg \tau^*$ viscous damping is negligible and the relaxation modulus is well approximated by its asymptote, $G_r \simeq G_0$. For the data in Fig.~\ref{fig:stressrelax}a the crossover time is $\tau^* \approx 10^{4}\tau_0$. In the following Section we will determine the scaling of $\tau^*$ with pressure.

\subsection{Scaling in the relaxation modulus} 
\label{sec:scaling}

\begin{figure}
 \includegraphics[width=1.0\columnwidth]{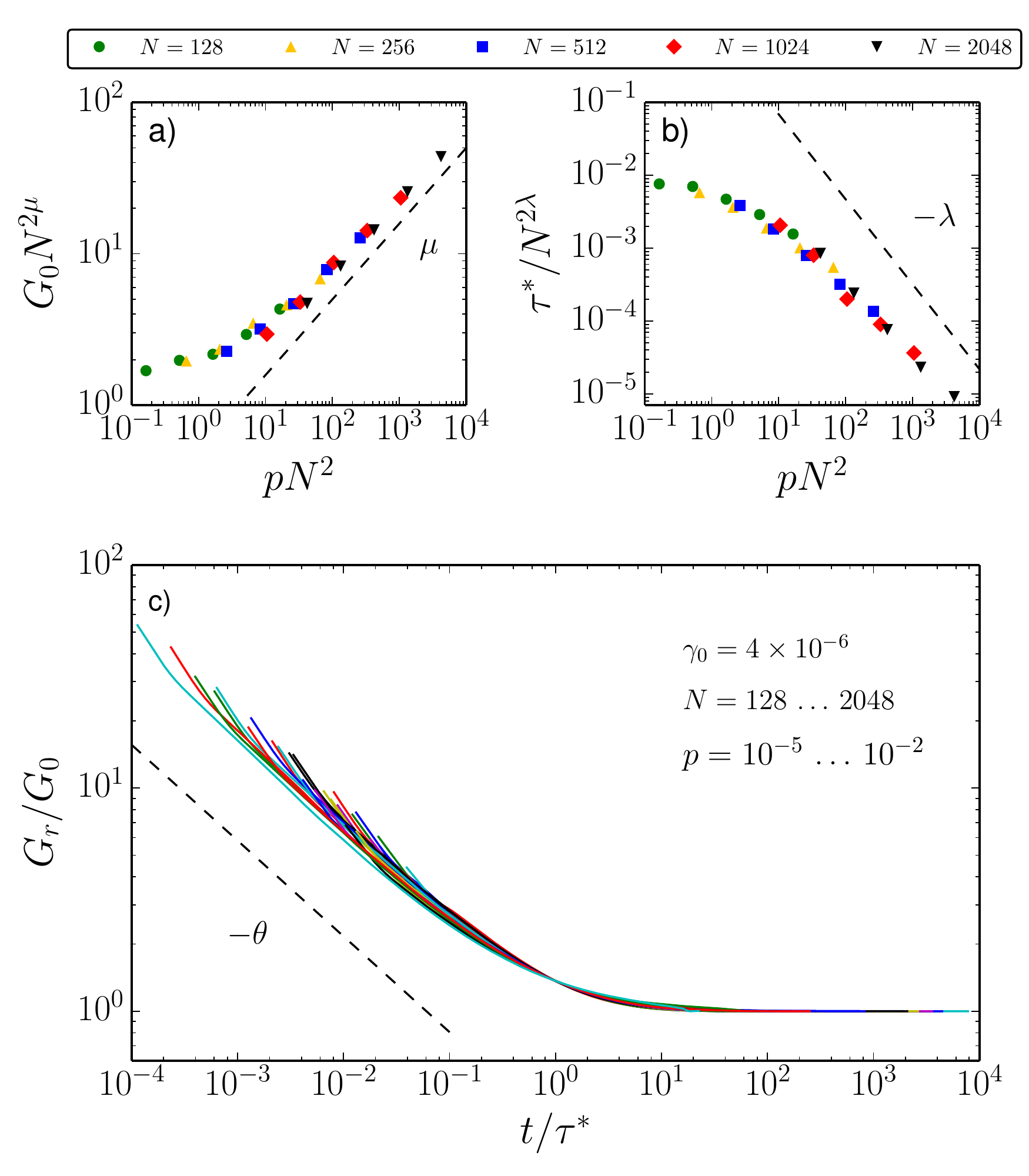}
\caption{(a) Finite size scaling collapse of the linear shear modulus $G_0$ in harmonic packings with exponent $\mu = 1/2$. 
(b) Finite size scaling collapse of the relaxation time $\tau^*$ with exponent $\lambda \approx 1.13$. 
(c) The relaxation modulus $G_r$ collapses to a master curve when $G_r$ and $t$ are  rescaled with $G_0$ and $\tau^*$, respectively, as determined in (a) and (b). At short times the master curve decays as a power law with exponent $\theta = \mu/\lambda \approx 0.44$ (dashed line).}
\label{fig:FS}
\end{figure}

We now characterize stress relaxation in linear response by measuring the relaxation modulus,  ensemble-averaged  over ensembles of packings prepared at varying pressure. We will show that $G_r$ collapses to a critical scaling function governed by the distance to the jamming point,
consistent with recent theoretical predictions by Tighe \cite{tighe11}.  
Our main focus is on numerically measuring the time scale beyond which viscous effects fade and the response becomes quasistatic,  which is predicted to scale as $\tau^* \sim {k \tau_0}/{p}$.

We showed in Fig.~\ref{fig:stressrelax} that a packing relaxes in three stages. The short-time plateau is trivial, in the sense that viscous forces prevent the particles from relaxing at rates faster than $1/\tau_0$; hence particles have not had time to depart significantly from the imposed affine deformation and the relaxation modulus reflects the contact stiffness, $G_r \sim k$. We therefore focus hereafter on the response on time scales $t \gg \tau_0$.

To demonstrate dynamic critical scaling in $G_r$, we first determine the scaling of its long-time asymptote $G_0$. We then identify the time scale $\tau^*$ on which $G_r$ significantly deviates from $G_0$. Finally, we show that rescaling with these two parameters collapses the relaxation moduli for a range of pressures to a single master curve. While we address variations with strain in subsequent Sections, the strain amplitude here is fixed to a value $\gamma_0 = 10^{-5.5}$. We have verified that this strain amplitude is in the linear regime for all of the data presented in this Section.

As noted above, at long times the relaxation modulus approaches the  linear quasistatic modulus, $G_r(t \rightarrow \infty) \simeq G_0$. 
We verify the scaling for $G_0$ from Eq.~(\ref{eqn:G0}) in our harmonic packings by repeating the finite size scaling analysis of Goodrich et al.~\cite{goodrich12}, who showed that finite size effects become important when a packing has $O(1)$ contacts in excess of isostaticity, or equivalently when $p/k \sim 1/N^2$ -- c.f.~Eq.~(\ref{eqn:order}). 
Consistent with their results, we find that ${\cal G} \equiv G_0 N^{2\mu}$ for varying $N$ and $p$ collapses to a master curve  when plotted versus $x \equiv pN^2$, as shown in Fig.~\ref{fig:FS}a. The scaling of Eq.~(\ref{eqn:G0}) is verified by this data collapse together with the requirement for the modulus to be an intensive property of large systems. To see this, note that $G_0$ is intensive only if ${\cal G} \sim x^{\mu}$ for large $x$.

Again referring to Fig.~\ref{fig:stressrelax}, there is clearly some time scale $\tau^*$ such that for $t < \tau^*$ the relaxation modulus deviates significantly from the quasistatic modulus.
To determine the scaling of $\tau^*$ with $p$, we perform the finite size scaling analysis presented in Fig.~\ref{fig:FS}b. 
The relaxation time is determined from the point where $G_r$, averaged over an ensemble of  at least 100 packings per condition, has decayed to within a fraction $\Delta$ of its final value,  $G_r(t = \tau^*) = (1+\Delta)G_0$. We present data for $\Delta = 1/e$, but similar scaling results for a range of $\Delta$ \cite{dagois-bohy14}. 
We require the rescaled pressure to remain $x = pN^2$ and collapse the data by rescaling the relaxation time as $\tau^* / N^{2\lambda}$ for a positive exponent $\lambda$.  It follows that  $\tau^*$ diverges in large systems near jamming as
\begin{equation}
{\tau^*} \sim \left(\frac{k}{p}\right)^\lambda \tau_0 \,\,\,{\rm as}\,\,\, N \rightarrow \infty \,.
\end{equation}
We find the best data collapse for $\lambda = 1.13$, close to but somewhat higher than the value $\lambda = 1$ predicted by theory \cite{tighe11},  although our current numerical results do not exclude this possibility.

We now use the linear quasistatic modulus $G_0$ and the characteristic time scale $\tau^*$ to collapse the relaxation modulus to a master curve ${\cal R}(s)$. Fig.~\ref{fig:FS}c plots $ {\cal R} \equiv G_r/G_0$ versus $s \equiv t/\tau^*$ for a range of pressures and system sizes; data from the trivial affine regime at times $t < 10\tau_0$ have been excluded. The resulting data collapse is excellent, and the master curve it reveals has two scaling regimes: ${\cal R} \simeq 1$ for $s \gg 1$, and ${\cal R} \sim s^{-\theta}$ for $s \ll 1$.
The plateau at large $s$ occurs by construction and corresponds to the quasistatic scaling $G_r \simeq G_0$. 
The power law relaxation at shorter times corresponds to $G_r \sim G_0(t/\tau^*)^{-\theta}$ for some exponent $\theta$. By considering a marginal solid prepared at the jamming point, one finds that the prefactor of $t^{-\theta}$ cannot depend on the pressure. Invoking the pressure scaling of $G_0$ and $\tau^*$ in the large $N$ limit, identified above, we conclude that $\theta = \mu/\lambda$. Hence in large systems the relaxation modulus scales as
\begin{equation}
\frac{G_r(t)}{k} \sim 
\left \lbrace
\begin{array}{cc}
 \left({\tau_0}/{t} \right)^{\theta} & 1 \ll t/\tau_0 \ll ({k}/p)^\lambda \\
(p/k)^{\mu} 				 & ({k}/p)^\lambda  \ll t/\tau_0  \,.
\end{array} \right.
\label{eqn:Gr}
\end{equation}
with $\mu = 1/2$, $\lambda \approx 1.13$, and $\theta = \mu/\lambda \approx 0.44$.

Anomalous stress relaxation with exponent $\theta \approx 1/2$ was first observed in simulations below jamming \cite{hatano09} and is also found in disordered spring networks \cite{tighe12,sheinman12}.
It is  related via Fourier transform to the anomalous scaling of the frequency dependent complex shear modulus $G^* \sim (\imath \omega)^{1-\theta}$ found in viscoelastic solids near jamming \cite{tighe11}.
We revisit the scaling relation of Eq.~(\ref{eqn:Gr}) in Section \ref{sec:rate}.

\section{Finite strain}

\label{sec:QS}
\begin{figure}
  \includegraphics[width=\columnwidth]{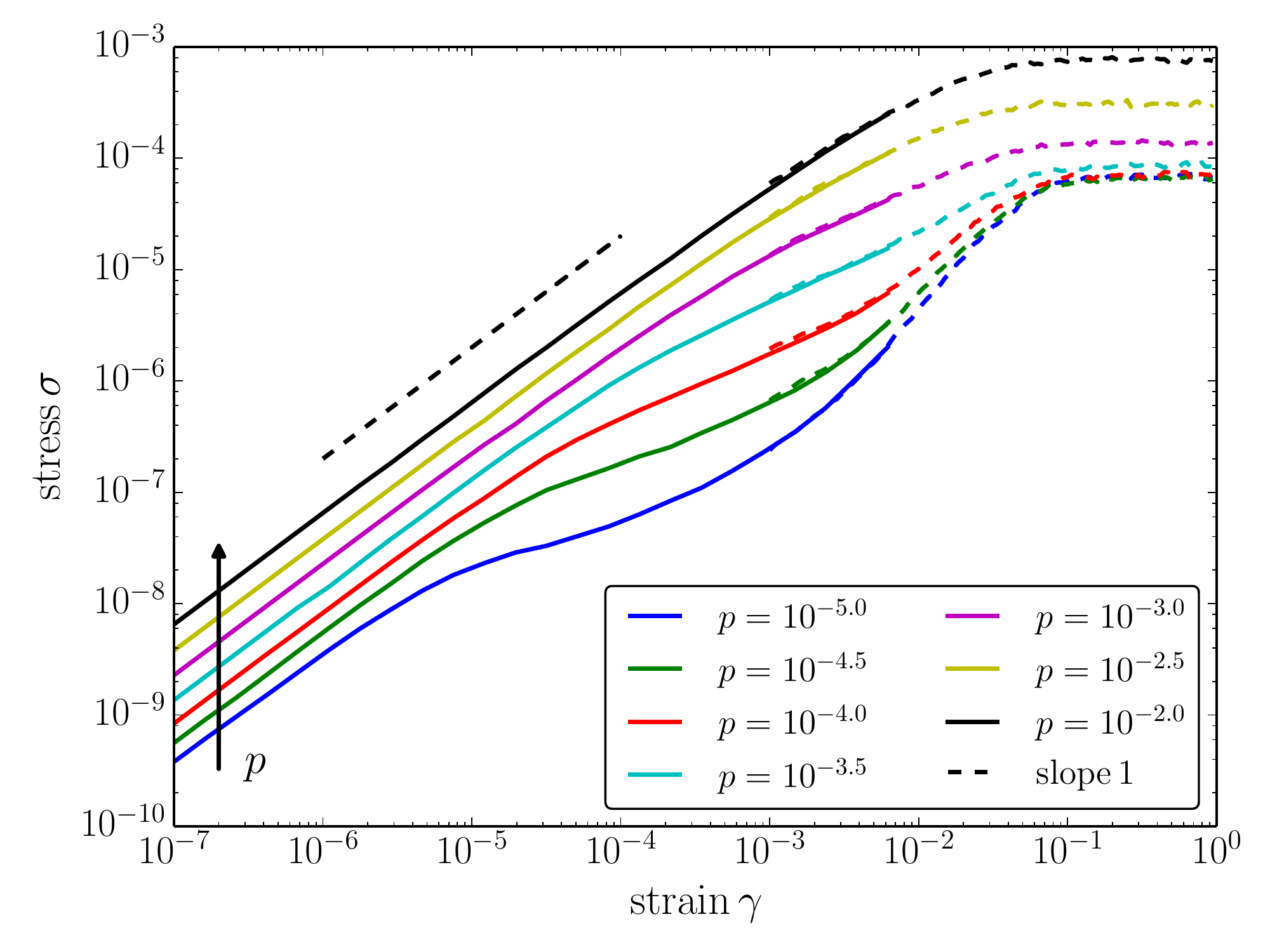}
  \caption{Averaged stress-strain curves under quasistatic shear at varying pressure $p$. Solid and dashed curves were calculated using different strain protocols.
Dashed curves: fixed strain steps of $10^{-3}$, sheared to a final strain of unity. 
Solid curves: logarithmically increasing strain steps, beginning at $10^{-9}$ and reaching a total strain of $10^{-2}$ after 600 steps.
  }
  \label{fig:plasticflow}
\end{figure}

When does linear elasticity break down under increasing strain, and what lies beyond?
To answer these questions, we now probe shear response at finite strain using flow start-up tests.

\subsection{Flow start-up}
In a flow start-up test, strain-controlled boundary conditions are used to ``turn on'' a flow with constant strain rate $\dot \gamma_0$ at time $t = 0$, i.e.
\begin{equation}
\gamma(t) = 
\left \lbrace
\begin{array}{cl}
0 & t< 0 \\
\dot\gamma_0 t & t \ge 0
\end{array} \right.
\label{eqn:startup}
\end{equation} 
To implement flow start-up in MD, at time $t =0$ a packing's particles and simulation cell are instantaneously assigned an affine velocity profile $\vec v_i = (\dot \gamma_0 \, y_i,0)^T$ in accordance with a simple shear with strain rate $\dot \gamma_0$; the Lees-Edwards images of the simulation cell are assigned a commensurate velocity. Then the particles are allowed to evolve according to Newton's laws while the Lees-Edwards boundary conditions maintain constant velocity, so that the total strain $\gamma(t)$ grows linearly in time. 

We also perform quasistatic shear simulations using nonlinear CG minimization to realize the limit of vanishing strain rate. Particle positions are evolved by giving the Lees-Edwards boundary conditions a series of small strain increments and equilibrating to a new minimum of the elastic potential energy. The stress $\sigma$ is then reported as a function of the accumulated strain. For some runs we use a variable step size in order to more accurately determine the response at small strain.

 Fig.~\ref{fig:window} illustrates the output of both the finite strain rate and quasistatic protocols.

\subsection{Quasistatic stress-strain curves}
To avoid complications due to rate-dependence, we consider the limit of vanishing strain rate first.

Fig.~\ref{fig:plasticflow} plots the ensemble-averaged stress-strain curve $\sigma(\gamma)$ for harmonic packings at varying pressure. Packings contain $N = 1024$ particles, and each data point is averaged over at least 600 configurations.  
Several features of the stress-strain curves stand out. First, there is indeed a window of initially linear growth. Second, beyond a strain of approximately 5 - 10\% the system achieves steady plastic flow and the stress-strain curve is flat. Finally, the end of linear elasticity and the beginning of steady plastic flow do not generally coincide; instead there is an interval in which the stress-strain curve has a complex nonlinear form. We shall refer to the end of the linear elastic regime as ``softening'' because the stress initially dips {\em below} the extrapolation of Hooke's law. (In the plasticity literature the same phenomenon would be denoted ``strain hardening''.) Moreover, for sufficiently low pressures there is a strain interval over which the stress increases faster than linearly. This surprising behavior is worthy of further attention, but the focus of the present work will be on the end of linear elasticity and the onset of softening. This occurs on a strain scale $\gamma^\dag$ that clearly depends on pressure.

\subsection{Onset of softening}

\begin{figure}
  \includegraphics[width=\columnwidth]{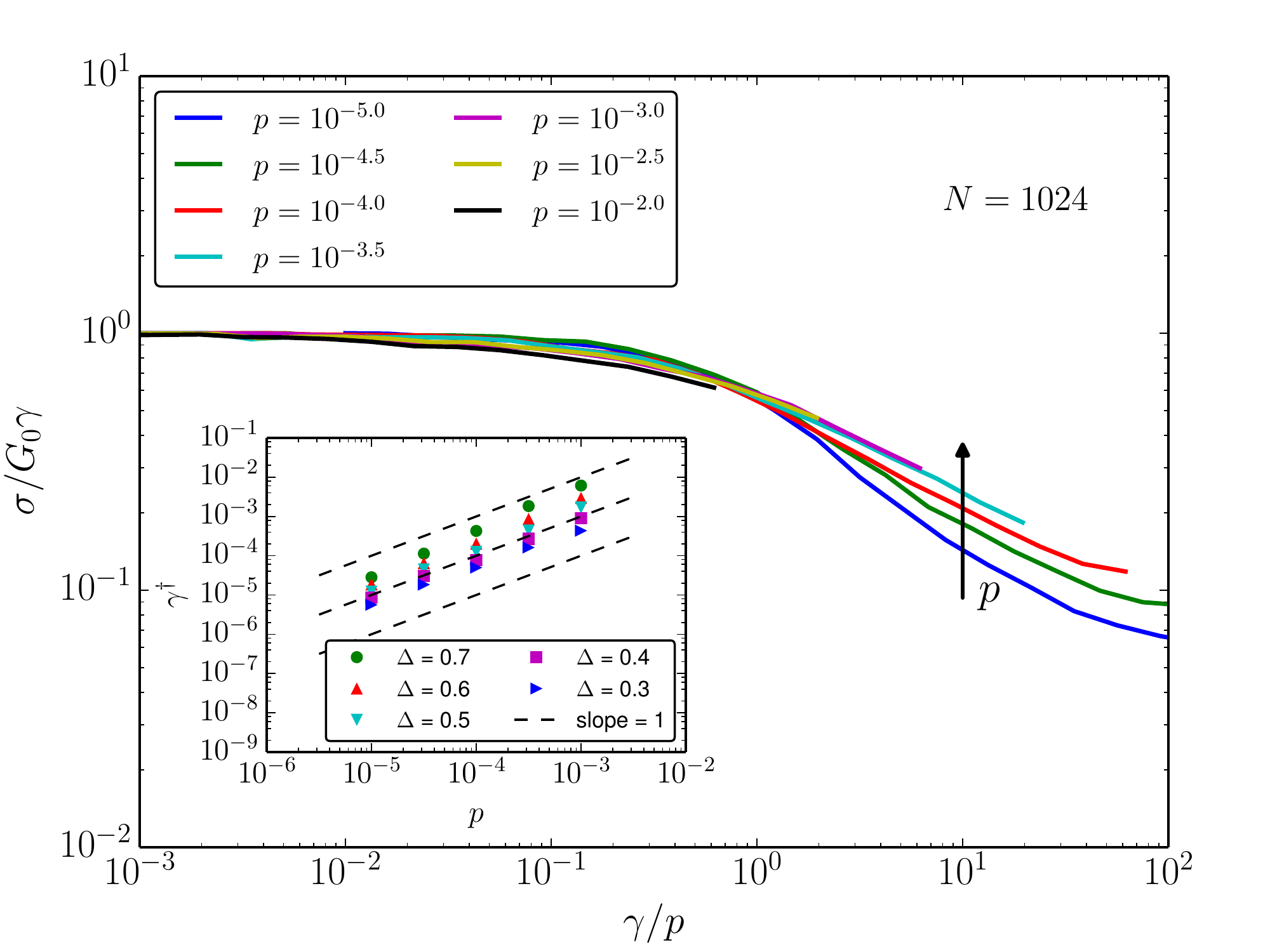}
  \caption{(main panel) Data from Fig.~\ref{fig:plasticflow}, expressed as a dimensionless effective shear modulus $\sigma/G_0 \gamma $ and plotted versus the rescaled strain $\gamma/p$.
(inset) The crossover strain $\gamma^\dag$ where the effective shear modulus has decayed by an amount $\Delta$ in a system of $N = 1024$ particles.
  }
\label{fig:softening}
\end{figure}

We now determine the pressure and system size dependence of the softening (or nonlinear) strain scale $\gamma^\dag$. 

Fig.~\ref{fig:softening} replots the quasistatic shear data from Fig.~\ref{fig:plasticflow} (solid curves), now with the linear elastic trend $G_0 \gamma$ scaled out. 
The rescaling collapses data for varying pressures in the linear regime and renders the linear regime flat. The strain axis in Fig.~\ref{fig:softening}b is also  rescaled with the pressure, a choice that will be justified below. The onset of softening occurs near unity in the rescaled strain coordinate for all pressures, which suggests that $\gamma^\dag$ scales linearly with $p$ in harmonic packings ($\alpha = 2$).

Unlike the linear relaxation modulus in Fig.~\ref{fig:FS}c, the quasistatic shear data in Fig.~\ref{fig:softening} do not collapse to a master curve; instead the slope immediately after softening steepens (in a log-log plot) as the pressure decreases. As a result, it is not possible to unambiguously identify a correlation $\gamma^\dag \sim p^\nu$ between the crossover strain and the pressure. To clarify this point, the inset of Fig.~\ref{fig:softening} plots the strain where $\sigma/G_0 \gamma$ has decayed by an amount $\Delta$ from its plateau value, denoted $\gamma^\dag(\Delta)$. This strain scale is indeed approximately linear in the pressure $p$ (dashed curves), but a power law fit gives an exponent $\nu$ in the range 0.87 to  1.06, depending on the value of $\Delta$. 
Bearing the above subtlety in mind, we nevertheless conclude that an effective power law with $\nu = 1$ provides a reasonable description of the softening strain. Section \ref{sec:scaling} presents further evidence to support this conclusion.

\subsection{Hertzian packings} 
In the previous section the pressure-dependence of $\gamma^\dag$ was determined for harmonic packings. We now generalize this result to other pair potentials, with numerical verification for the case of Hertzian packings ($\alpha = 5/2$). 

\begin{figure}
  \includegraphics[width=\columnwidth]{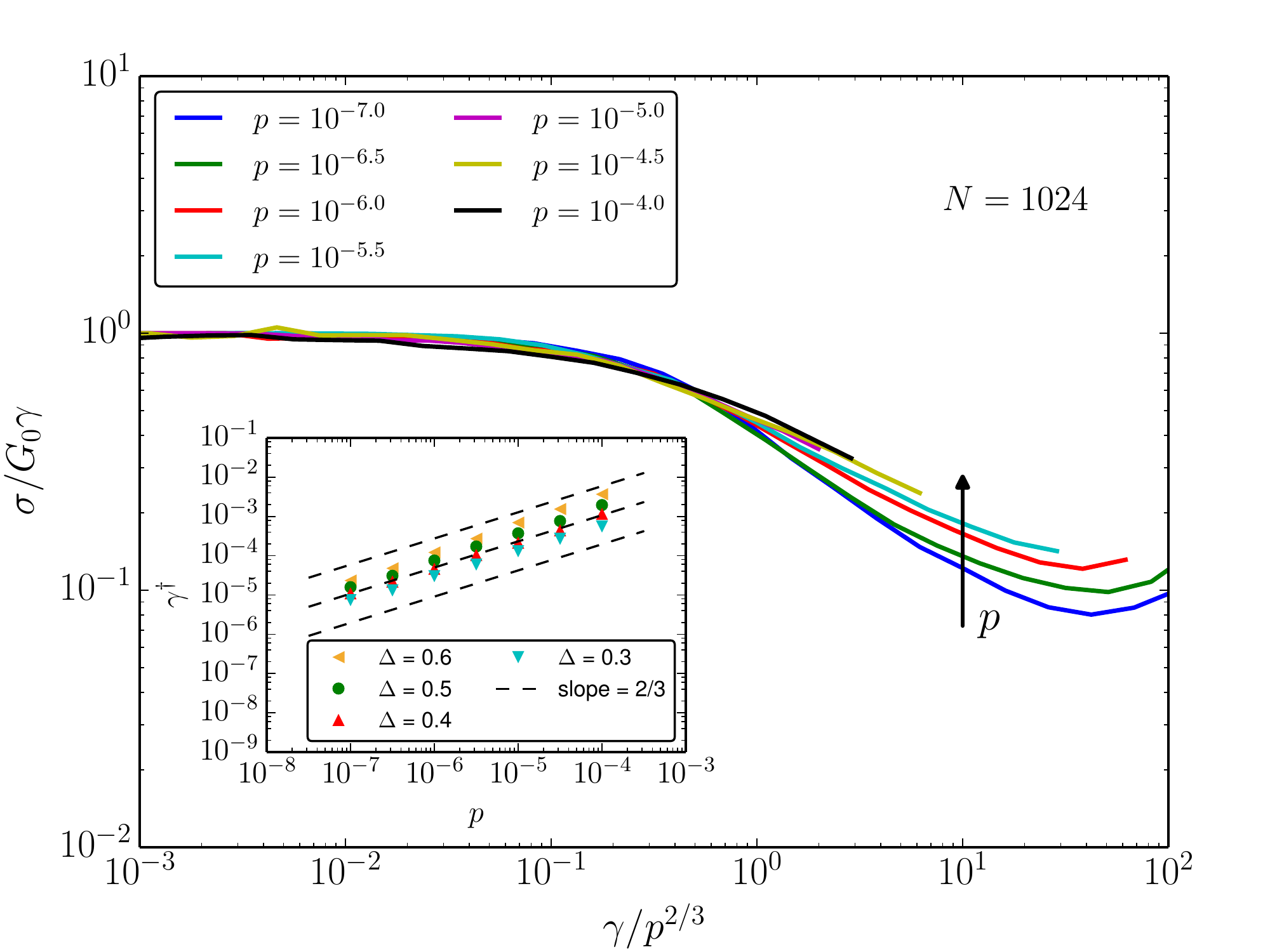}
\caption{(main panel) The dimensionless shear modulus of quasistatically sheared Hertzian packings plotted versus the rescaled strain $\gamma/p^{2/3}$. (inset) Pressure-dependence of the crossover strain $\gamma^\dag$. 
}
\label{fig:hertzian}
\end{figure}

Recall that the natural units of stress are set by the contact stiffness $k$, which itself varies with pressure when $\alpha \neq 2$. Based on the linear scaling of $\gamma^\dag$ in harmonic packings, we anticipate
\begin{equation}
\gamma^\dag \sim  \frac{p}{k}  \sim p^{1/(\alpha - 1)} \,,
\label{eqn:gdag}
\end{equation}
which becomes $\gamma^\dag  \sim p^{2/3}$ in the Hertzian case.
To test this relation, we repeat the analysis of the preceding Section; results are shown in Fig.~\ref{fig:hertzian}. We again find a finite linear elastic window that gives way to softening. Softening onset can again be described with a $\Delta$-dependent exponent (see inset). Its value has a narrow spread about $2/3$; power law fits give slopes between 0.63 and 0.74.

\subsection{Relating softening and contact changes}
\label{sec:cc}

\begin{figure}
  \includegraphics[width=\columnwidth]{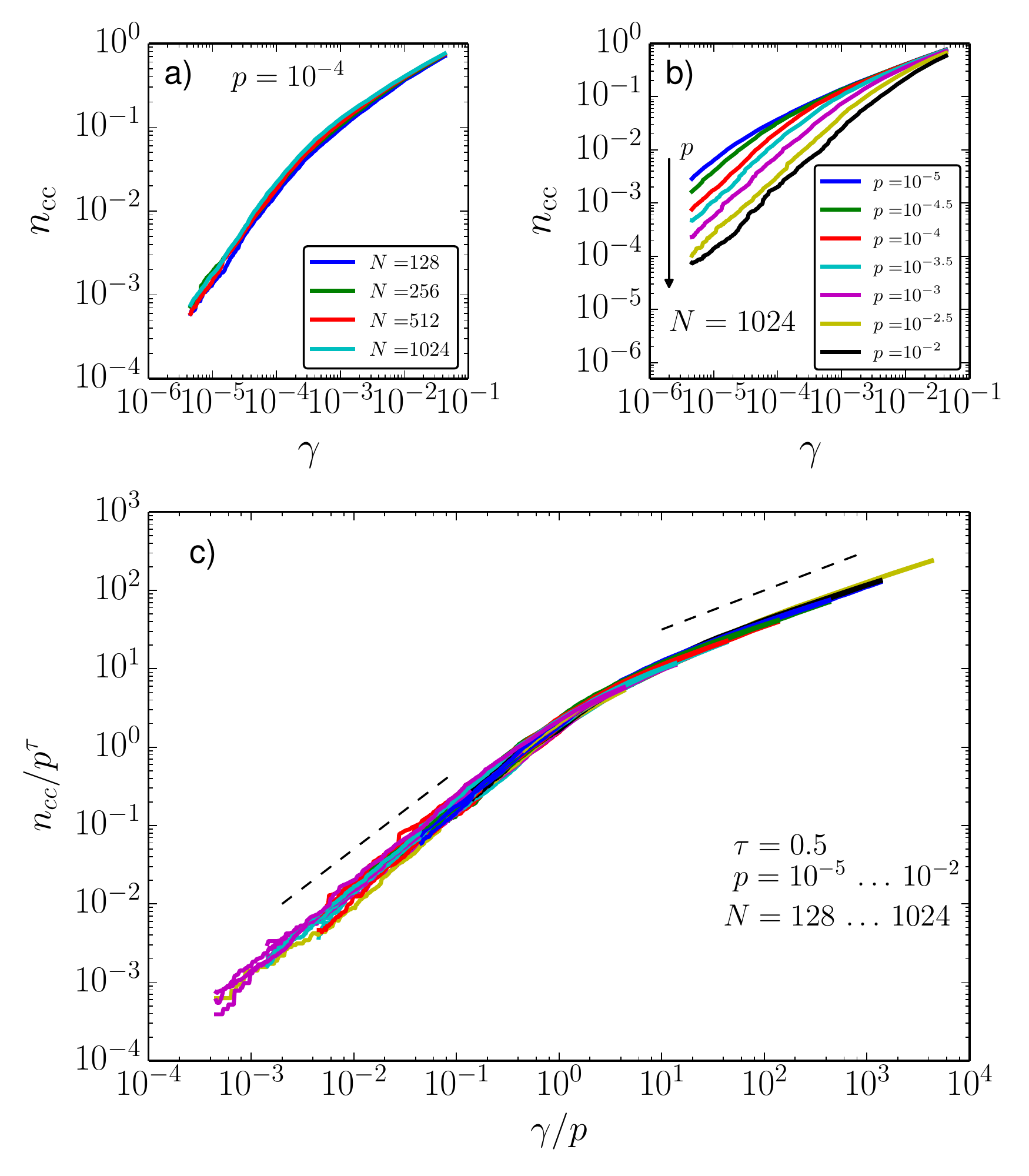}
\caption{The contact change density shown for (a) varying system size and (b) varying pressure. 
(c) Data collapse for pressures $p = 10^{-2} \ldots 10^{-5}$ in half decade steps and system sizes $N = 128 \ldots 1024$ in multiples of 2. Dashed lines indicate slopes of 1 and 1/2. 
}
\label{fig:Ncc}
\end{figure}
Why does the linear elastic window close when it does? We now seek to relate softening with contact changes on the particle scale \cite{schreck11,vandeen14,knowlton14,keim14,keim15,kawasaki15}. Specifically, we identify a correlation between the softening strain $\gamma^\dag$, the cumulative number of contact changes, and the distance to the isostatic contact number $z_c$. In so doing we will answer the question first posed by Schreck and co-workers \cite{schreck11}, who asked how many contact changes a packing can accumulate while still displaying linear elastic response.

We begin by investigating the ensemble-averaged contact change density $n_{\rm cc}(\gamma) \equiv [N_{\rm make}(\gamma) + N_{\rm break}(\gamma)]/N$, where $N_{\rm make}$  and $N_{\rm break}$ are the number of made and broken contacts, respectively, accumulated during a strain $\gamma$. Contact changes are identified by comparing the contact network at strain $\gamma$ to the network at zero strain.

In Fig.~\ref{fig:Ncc}a we plot $n_{\rm cc}$ for packings of harmonic particles at pressure $p = 10^{-4}$ and varying system size. The data collapse to a single curve, indicating that $n_{\rm cc}$ is indeed an intensive quantity.
The effect of varying pressure is shown in Fig.~\ref{fig:Ncc}b.  There are two qualitatively distinct regimes in $n_{\rm cc}$, with a crossover governed by pressure.

To better understand these features, we seek to collapse the $n_{\rm cc}$ data  to a master curve.  
By plotting ${\cal N} \equiv n_{\rm cc}/p^{\tau}$ versus $y \equiv \gamma/p$, we obtain excellent collapse for $\tau = 1/2$, as  shown in Fig.~\ref{fig:Ncc}b for the same pressures as in Fig.~\ref{fig:Ncc}a and system sizes $N = 128 \ldots 1024$. The scaling function ${\cal N} \sim y$ for small $y$, while ${\cal N} \sim y^\tau$ for $y \gtrsim 1$. The rescaled strain $y$ provides further evidence for a crossover scale $\gamma^\dag \sim p/k$, now apparent at the microscale. Moreover, the fact that data for varying system sizes all collapse to the same master curve is an important indicator that $\gamma^\dag$ is an intensive strain scale that remains finite in the large system size limit.

The scaling collapse in Fig.~\ref{fig:Ncc}c generalizes the results of Van Deen et al.~\cite{vandeen14}, who determined the strain scale $\gamma_{\rm cc}^{(1)} \sim (p/k)^{1/2}/N$ associated with the first contact change. To see this, note that  the inverse slope $({\rm d}\gamma/{\rm d}n_{\rm cc})/N$ represents the average strain interval between contact changes at a given strain. Hence the initial slope of $n_{\rm cc}$ is fixed by $\gamma_{\rm cc}^{(1)}$:
\begin{equation}
n_{\rm cc}(\gamma) \simeq \frac{1}{N} \left(\frac{\gamma}{\gamma_{\rm cc}^{(1)}} \right)  \,\,\,\,\,\, {\rm as } \,\,\,\,\,\, \gamma \rightarrow 0  \,. 
\label{eqn:ncc}
\end{equation}
From Fig.~\ref{fig:Ncc} it is apparent that $n_{\rm cc}$ remains linear in $\gamma$ up to the crossover strain $\gamma^\dag$. We conclude that $\gamma_{\rm cc}^{(1)}$ describes the strain between successive contact changes over the entire interval $0 \le \gamma < \gamma^\dag$. In the softening regime the strain between contact changes increases; it scales as $n_{\rm cc} \sim \gamma^{1/2}$ (see Fig.~\ref{fig:Ncc}c).

Let us now re-interpret the softening crossover strain $\gamma^\dag \sim \Delta z^2$ (c.f.~Eq.~(\ref{eqn:order})) in terms of the coordination of the contact network. 
We recall that $\Delta z = z - z_c$ is the difference between the initial contact number $z$ and the isostatic value $z_c$, which corresponds to the minimum number of contacts per particle needed for rigidity. The excess coordination $\Delta z$ is therefore an important characterization of the contact network.
The contact change density at the softening crossover, $n_{\rm cc}^\dag$, can be related to $\Delta z$ via Eq.~(\ref{eqn:ncc}), while making use of Eq.~(\ref{eqn:order}),
\begin{equation}
n_{\rm cc}^\dag \equiv n_{\rm cc}(\gamma^\dag) \sim \Delta z\,.
\end{equation}
Hence we have empirically identified a topological criterion for the onset of softening: an initially isotropic packing softens when it has undergone an extensive number of contact changes that is comparable to the number of contacts it initially had in excess of isostaticity. (This does not mean the packing is isostatic at the softening crossover, as $n_{\rm cc}$ counts both made and broken contacts.)

\subsection{Rate-dependence}
\label{sec:rate}
To this point we have considered nonlinear response exclusively in the limit of quasistatic shearing.
A material accumulates strain quasistatically when the imposed strain rate is slower than the longest relaxation time in the system.
Because relaxation times near jamming are long and deformations in the lab always occur at finite rate, we can anticipate that quasistatic response is difficult to achieve and that rate-dependence generically plays a significant role. Hence it is important to consider shear at finite strain and finite strain rate. We now consider flow start-up experiments in which a finite strain rate $\dot \gamma_0$ is imposed at time $t = 0$, cf.~Eq.~(\ref{eqn:startup}). 

\begin{figure}
  \includegraphics[width=\columnwidth]{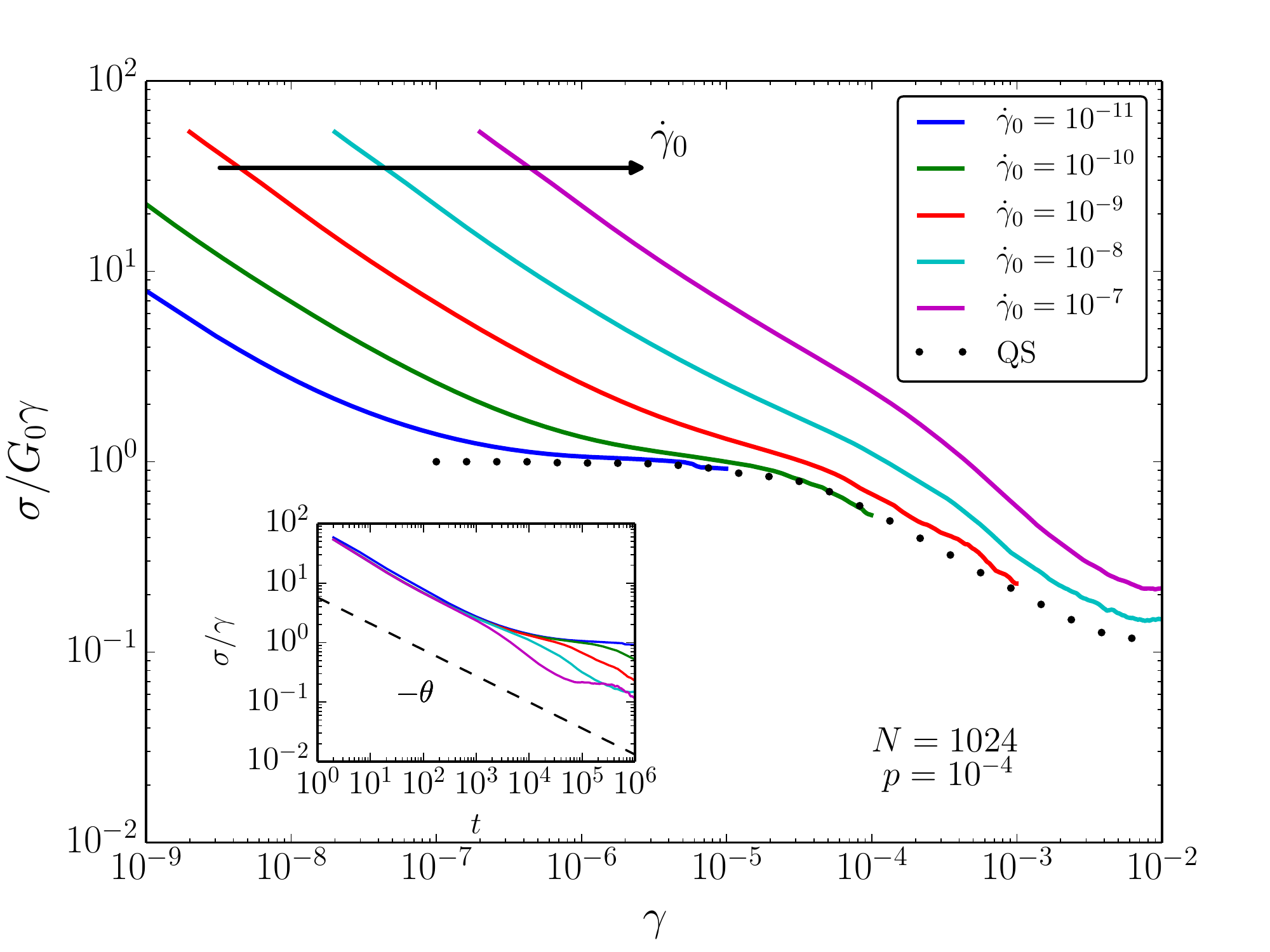}
\caption{The effective shear modulus during flow start-up for packings of $N = 1024$ particles at pressure $p = 10^{-4}$, plotted versus strain for varying strain rates $\dot \gamma_0$.  (inset) The same data collapses for early times when plotted versus $t$, decaying as a power law with exponent $\theta = \mu/\lambda \approx 0.44$ (dashed line). }
\label{fig:flowstartup}
\end{figure}

Fig.~\ref{fig:flowstartup} displays the mechanical response to flow start-up for varying strain rates. To facilitate comparison with the quasistatic data of the previous section, flow start-up data are plotted in terms of the dimensionless quantity $ \sigma(t;\dot \gamma_0)/G_0 \gamma$, which we shall refer to as the effective shear modulus.  The data are for systems of $N = 1024$ particles, averaged over an ensemble of around 100 realizations each.  Here we plot data for the pressure $p = 10^{-4}$; results are qualitatively similar for other pressures.  For comparison, we also plot the result of quasistatic shear (solid circles) applied to the same ensemble of packings.

Packings sheared sufficiently slowly follow the quasistatic curve; see e.g.~data for $\dot \gamma_0 = 10^{-11}$. 
For smaller strains, however, the effective shear modulus is stiffer than the quasistatic curve and decays as $\sigma/\gamma \sim t^{-\theta}$ (see inset). This is rate-dependence: for a given strain amplitude, the modulus increases with increasing strain rate. Correspondingly, the characteristic strain $\gamma^*$ where curves in the main panel of Fig.~\ref{fig:flowstartup} reach the linear elastic plateau ($\sigma/G_0 \gamma \approx 1$) grows with $\dot \gamma_0$.
For sufficiently high strain rates there is no linear elastic plateau; for the data in Fig.~\ref{fig:flowstartup} this occurs for $\dot \gamma_0 \approx 10^{-8}$. Hence there is a characteristic strain rate, $\dot \gamma^\dag$, beyond which the linear elastic window has closed: packings sheared faster than $\dot \gamma^\dag$ are always rate-dependent and/or strain softening.

To understand the rate-dependent response at small strains, we revisit the relaxation modulus determined in Section \ref{sec:relax}.  In linear response the stress after flow start-up depends only on the elapsed time $t  = \gamma / \dot \gamma_0$,
\begin{equation}
\frac{\sigma}{ \gamma} = \frac{1}{t} \, \int_0^{t} G_r(t') \, {\rm d}t' \,.
\end{equation}
Employing the scaling relations of Eq.~(\ref{eqn:Gr}), one finds  
\begin{equation}
\frac{\sigma}{ \gamma}  \sim  k \left(\frac{ \tau_0}{t}\right)^{\theta}, \,\,\,\,\,\,\,\,\,\,\,\, \tau_0 < t <  
\tau^* \,,
\end{equation}
as verified in Fig.~\ref{fig:flowstartup} (inset).
Linear elasticity ${\sigma}/{ \gamma}  \simeq G_0$ is only established at longer times, when $\gamma > \dot \gamma_0 \tau^*  \sim ({k}/{p})^\lambda\,\dot \gamma_0 \tau_0$.
Hence the relaxation time $\tau^*$ plays an important role: it governs the crossover from rate-dependent to quasistatic linear response. The system requires a time $\tau^*$ to relax after a perturbation. When it is driven at a faster rate, it cannot relax fully and hence its response depends on the driving rate.

We can now identify the characteristic strain rate $\dot \gamma^\dag$ where the linear elastic window closes.
This  rate is reached when the bound on quasistaticity, $\gamma > \dot \gamma_0 \tau^*$, collides with the bound on linearity, $\gamma < \gamma^\dag$, giving
\begin{equation}
\dot \gamma^\dag \sim \frac{(p/k)^{1+\lambda}}{\tau_0} \,,
\end{equation} 
with $1+\lambda \approx 2.1$.
This strain rate vanishes rapidly near jamming, and packings must be sheared increasingly slowly to observe a stress-strain curve that obeys Hooke's law. As a practical consequence, experiments near jamming are unlikely to access the linear elastic regime.

\begin{figure}
  \includegraphics[width= 0.7\columnwidth]{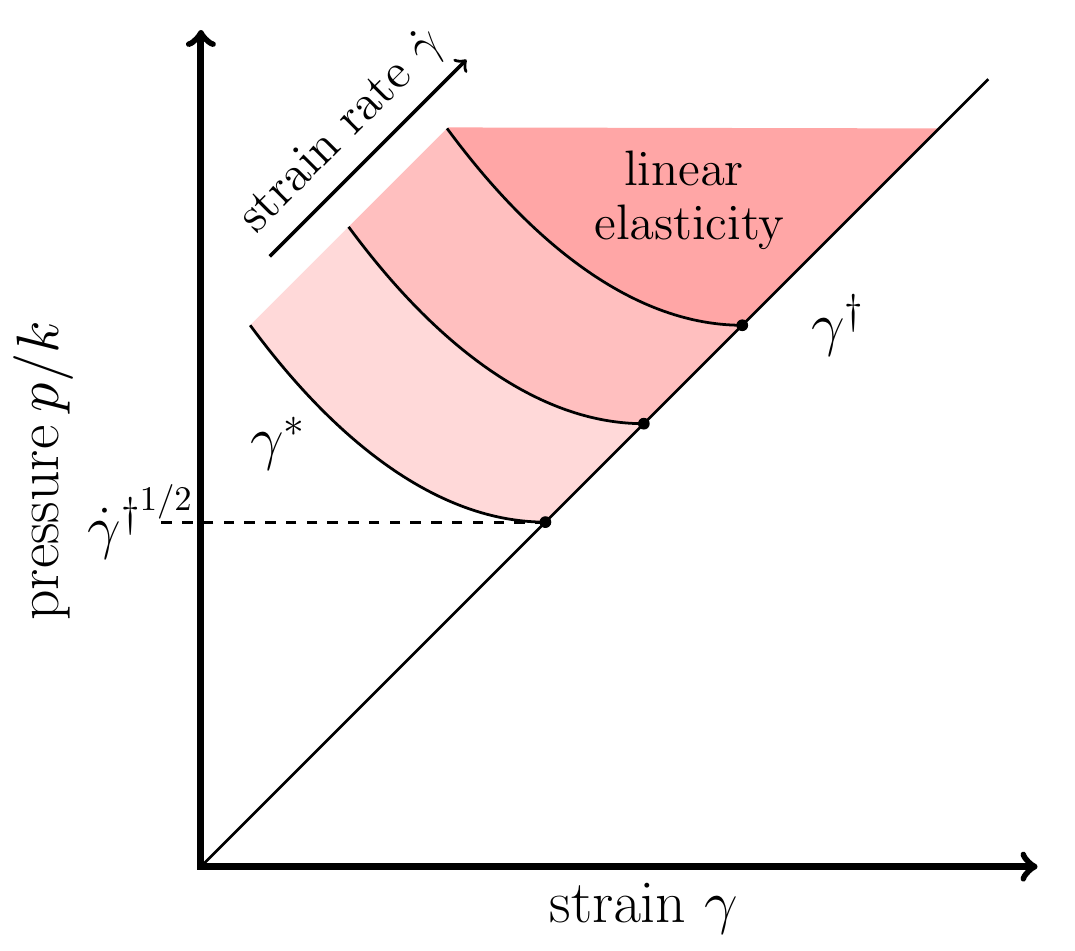}
\caption{In a flow start-up test, quasistatic linear response ($G \approx G_0$) occupies a strain window $\gamma^* < \gamma < \gamma^\dag$ (shaded regions). For smaller strains the response is rate-dependent, with a crossover strain $\gamma^*$ that depends on both pressure and strain rate. Softening sets in for higher strains, with a crossover $\gamma^\dag$ that depends only on the pressure. The intersection of the rate-dependent and softening crossovers defines a strain rate $\dot \gamma^\dag$ above which there is no quasistatic linear response, i.e.~the shaded region closes. }
\label{fig:regimes}
\end{figure}

\section{Discussion}

Using a combination of stress relaxation and flow start-up experiments, we have shown that soft solids near jamming are easily driven out of the linear elastic regime. There is, however, a narrow linear elastic window that survives the accumulation of an extensive number of contact changes. This window is bounded from below by rate-dependent viscoelasticity and bounded from above by the onset of strain softening. Close to the transition these two bounds collide and the linear elastic window closes. Finally, weakly jammed materials are generally rate-dependent and/or strain softening on scales relevant to the laboratory, because the strains and strain rates bounding the linear elastic window vanish rapidly near jamming.
Fig.~\ref{fig:regimes} provides a qualitative summary of our results. 

While our simulations are in two dimensions, we expect the scaling relations we have identified to hold for $D>2$. To the best of our knowledge, all scaling exponents near jamming that have been measured in both 2D and 3D are the same. There is also numerical evidence that $D = 2$ is the transition's upper critical dimension \cite{goodrich12,goodrich14}.

Our work provides a bridge between linear elasticity near jamming, viscoelasticity at finite strain rate, and nonlinearity at finite strain amplitude. The measured relaxation modulus $G_r$ is in good agreement with the linear viscoelasticity predicted by Tighe \cite{tighe11}. Consistent with the granular experiments of Coulais et al., we identify a crossover to nonlinear strain softening. Their crossover scales differently with the distance to jamming, possibly due to the presence of static friction. The emulsions of Knowlton et al.~also soften \cite{knowlton14}. They display a crossover strain that is roughly linear in $\Delta \phi$, consistent with both our $\gamma^\dag$ and the results of Otsuki and Hayakawa \cite{otsuki14}, who simulated large amplitude oscillatory shear at finite frequency. 
The agreement between the crossover strains in our quasistatic simulations and the oscillatory shear simulations of Ref.~\cite{otsuki14} is surprising, as most of their results are for frequencies higher than $\dot \gamma^\dag$, where viscous stresses dominate.
There are also qualitative differences between the quasistatic shear modulus, which cannot be collapsed to a master curve (Fig.~\ref{fig:softening}), and the storage modulus in oscillatory shear, which can \cite{otsuki14,dagois-bohy14}. We speculate that there are corresponding microstructural differences between packings in steady state and transient shear \cite{regev13}, similar to those which produce memory effects \cite{keim13}.

Soft sphere packings near jamming approach the isostatic state, which also governs the rigidity of closely related materials such as biopolymer and fiber networks \cite{heussinger06,heussinger06b,broedersz11,das12}. It is therefore remarkable to note that, whereas sphere packings soften under strain, quasistatically sheared amorphous networks are strain stiffening beyond a crossover strain that scales as $\Delta z$ \cite{wyart08}, which vanishes more slowly than $\gamma^\dag \sim \Delta z^2$ in packings. Hence nonlinearity sets in later and with opposite effect in networks \cite{tighe14}. We expect that this difference is attributable to contact changes, which are absent or controlled by slow binding/unbinding processes in networks. 

We have demonstrated that the onset of softening occurs when the system has accumulated a finite number of contact changes correlated with the system's initial distance from the isostatic state. This establishes an important link between microscopic and bulk response. Yet further work investigating the relationship between microscopic irreversibility, softening, and yielding is needed. The inter-cycle diffusivity in oscillatory shear, for example, jumps at yielding \cite{knowlton14,kawasaki15}, but its pressure dependence has not been studied. Shear reversal tests could also provide insight into the connection between jamming and plasticity.

While the onset of softening can be probed with quasistatic simulation methods, rate dependent effects such as the strain scale $\gamma^*$ should be sensitive to the manner in which energy is dissipated. The dissipative contact forces considered here are most appropriate as a model for  foams and emulsions. Hence useful extensions to the present work might consider systems with, e.g., lubrication forces or a thermostat.

\section{Acknowledgments}
We thank P.~Boukany, D.~J.~Koeze, M.~van Hecke, and S.~Vasudevan for valuable discussions.
JB, DV and BPT were supported by the Dutch Organization for Scientific Research (NWO). ES was supported by the J\'anos Bolyai Research Scholarship of the Hungarian Academy of Sciences.
This work was carried out on the Dutch national e-infrastructure with the support of SURF Cooperative.

%

\end{document}